\documentclass[11pt,letterpaper]{article}
\pdfoutput=1
\usepackage{a4wide,epsfig,amsmath,amssymb,cite,scalefnt,graphicx}
\numberwithin{equation}{section}
\allowdisplaybreaks 
\usepackage{array}
\usepackage{epstopdf}
\usepackage{tikz}
\usetikzlibrary{intersections}
\usetikzlibrary{calc}
\usetikzlibrary{shapes}
\usetikzlibrary{arrows,shapes,positioning}
\usetikzlibrary{decorations.markings}
\tikzstyle arrowstyle=[scale=1]
\tikzstyle directed=[postaction={decorate,decoration={markings,
    mark=at position .5 with {\arrow[arrowstyle]{stealth}}}}]
\tikzstyle reverse directed=[postaction={decorate,decoration={markings,
    mark=at position .5 with {\arrowreversed[arrowstyle]{stealth};}}}]

\def\phibar{\bar\phi}
\def\be{\begin{equation}}
\def\ee{\end{equation}}
\def\pa{\partial}
\def\Ocal{{\cal O}}
\def\sigmabar{\overline\sigma}
\def\nn{\nonumber\\}
\def\Tr{\hbox{Tr}}
\def\Jcal{{\cal J}}
\def\atilde{\tilde a}
\def\abar{\overline a}
\def\mutilde{\tilde \mu}

\input{epsf}

\begin{document}

\begin{titlepage}
\begin{flushright}
LTH1264\\
{\today
}\\

\end{flushright}
\date{}
\vspace*{3mm}

\begin{center}
{\Huge Anomalous dimensions at large charge for $U(N)\times U(N)$ theory in three and four dimensions}\\[12mm]
{\bf I.~Jack\footnote{{\tt dij@liverpool.ac.uk}} and  D.R.T.~Jones\footnote{{\tt drtj@liverpool.ac.uk}} 
}\\

\vspace{5mm}
Dept. of Mathematical Sciences,
University of Liverpool, Liverpool L69 3BX, UK\\

\end{center}

\vspace{3mm}
\begin{abstract}
Recently it was shown that the scaling dimension of the operator $\phi^n$ in $\lambda(\phibar\phi)^2$ theory may be computed semiclassically at the Wilson-Fisher fixed point in $d=4-\epsilon$, for generic values of $\lambda n$, and this was verified to two loop order in perturbation theory at leading and subleading $n$. This result was subsequently generalised to operators of fixed charge $Q$ in $O(N)$ theory and verified up to four loops in perturbation theory at leading and subleading $Q$. More recently, similar semiclassical calculations have been performed for the classically scale-invariant $U(N)\times U(N)$ theory in four dimensions, and verified up to two loops, once again at leading and subleading $Q$. Here we extend this verification to four loops. We also consider the corresponding classically scale-invariant theory in three dimensions, similarly verifying the leading and subleading semiclassical results up to four loops in perturbation theory.
\end{abstract}

\vfill

\end{titlepage}

\section{Introduction}

The investigation  of 
renormalizable scalar field theories with scale invariant self-interactions has attracted renewed attention in recent years. Such theories have been central to the study of conformal field theory; in particular, the development of the theory of critical phenomena was illuminated by the study of quartic ($\phi^4$) interactions in $d=4-\epsilon$ dimensions in the crucial work of Wilson~\cite{wils1,wils2}\ and Wilson and Fisher~\cite{wf}. Critical indices for the theory are determined via the renormalisation constants of the theory.

In recent years, considerable progress~\cite{son,horm,alos,aos,rod,Bad2,sann,sann2,sann3,alos2} has been made for this class of theories, using a semi-classical expansion in the path integral formulation of the theory\footnote{A similar approach was also pursued for $\phi^6$ theories for $d=3-\epsilon$ and $\phi^3$ theories for $d=6-\epsilon$ in Refs.~\cite{Bad,rod2,JJ}. Note, however, that Ref.~\cite{rod2} does not perform any of the kind of perturbative analysis which has been featured in 3 and 4 dimensions.}. In particular, this approach gives a useful handle on $n$-point amplitudes for large $n$.

In Ref~\cite{rod} the anomalous dimension of the $\phi^n$ operator was considered in the  $O(N)$-invariant $g(\phi^2)^2$ theory with an $N$-dimensional scalar multiplet $\phi$, for large $n$ and  fixed $gn^2$. In Ref.~\cite{Bad2} the scaling dimension of the same operator in the $U(1)$-invariant $\lambda(\phibar\phi)^2$ theory (corresponding to the special case $N=2$) was computed at the Wilson-Fisher fixed point $\lambda_*$ as a semiclassical expansion in $\lambda_*$, for fixed $\lambda_*n$. Subsequently this was generalised in Ref.~\cite{sann} to the case of an operator of charge $Q$ in the $O(N)$-invariant  theory.  In Ref.~\cite{Bad2}, the $U(1)$ result was compared with perturbation theory up to two loops, and in Refs.~\cite{sann}, \cite{JJ1} the check was performed for the $O(N)$ theory up to three and four loops respectively. Most recently, the classically scale-invariant $U(N)\times U(N)$ model has been investigated in Ref.~\cite{sann2}. Here once again the leading and non-leading terms in a large charge expansion have been derived by a semi-classical calculation, and compared with perturbative results up to two loop order. In the current work we extend this check to three and four loops in perturbation theory, and further perform a similar analysis and check for the classically scale-invariant $U(N)\times U(N)$ theory in three dimensions.

The paper is organised as follows: In Section 2 we describe the semiclassical calculation in the $U(N)\times U(N)$ case, following Ref.~\cite{sann2}. Then in Section~3 we compare the result of this calculation with perturbative calculations up to and including 4 loops. This represents a significant extension of previous calculations. In Section~4 we extend these semiclassical results to the corresponding classically scale-invariant theory in three dimensions, and perform a perturbative check through four loops.

\section{The $U(N)\times U(N)$ model}

Our discussion in this section follows closely that of Ref.~\cite{sann2}, where the semiclassical calculation was first performed for the $U(N)\times U(N)$ model.
In this model we have an $n\times n$ complex matrix $H$, and the Lagrangian is given by
\be
{\cal L} =\Tr\left(\pa^{\mu}H^{\dagger}\pa_{\mu}H\right)+u\Tr\left(H^{\dagger}HH^{\dagger}H\right)+v\left[\Tr\left(H^{\dagger}H\right)\right]^2.
\label{lag4}
\ee
Note that one qualitative difference with earlier models considered in this context is that this is a multi-coupling theory. As usual we work within dimensional regularisation with $d=4-\epsilon$ and with the divergences for this theory cancelled by replacing the couplings by bare counterparts given to leading (one-loop) order by
\begin{align}
\alpha_y^{B(1)}=&\alpha_y+\frac{1}{\epsilon}\left[\frac{2}{N}\alpha_y^2(N^2+4)+16\alpha_y\alpha_h+\frac{24}{N}\alpha_h^2\right],\nn
\alpha_h^{B(1)}=&\alpha_h+\frac{1}{\epsilon}\left[\frac{12}{N}\alpha_y\alpha_h+8\alpha_h^2\right],
\label{bare}
\end{align}
where
\be
\alpha_h=\frac{uN}{16\pi^2},\quad \alpha_y=\frac{2vN}{16\pi^2}.
\ee
In principle of course the field $H$ is also replaced by a bare counterpart, but for this theory this only becomes necessary from two-loop order. For the purposes of the semiclassical calculation presented here, it is mostly sufficient to work with the renormalised couplings in order to avoid a proliferation of indices; we shall point out the single instance where we need to reinstate the bare couplings.  The $\beta$-functions are given by\cite{pisarski} 
\begin{align}
\beta_{\alpha_y}=&-\epsilon \alpha_y+\frac{2}{N}\alpha_y^2(N^2+4)+16\alpha_y\alpha_h+\frac{24}{N}\alpha_h^2+\ldots,\nn
\beta_{\alpha_h}=&-\epsilon \alpha_h+\frac{12}{N}\alpha_y\alpha_h+8\alpha_h^2+\ldots,
\label{betas}
\end{align}
where we display only the one-loop contributions corresponding to the one-loop bare couplings in Eq.~\eqref{bare}.
The conformal fixed point is defined by $\alpha_h=\alpha_h^*$, $\alpha_y=\alpha_y^*$, with  $\alpha_h^*$, $\alpha_y^*$ satisfying
\be
\beta_{\alpha_h}=\beta_{\alpha_y}=0.
\label{gfix}
\ee
We do not need to give $\alpha_h^*$, $\alpha_y^*$ explicitly, but as pointed out in Ref.~\cite{sann2}, the fixed point is complex for $N>\sqrt3$, which may be associated with ``walking'' behaviour of the couplings near the fixed point.

We initially work in general $d$. The semiclassical computation is performed by mapping the theory at the conformal fixed point via a Weyl transformation to a cylinder $\mathbb{R}\times S^{d-1}$, where $S^{d-1}$ is a sphere of radius $R$, giving
\begin{align}
S_{\rm{cyl}}=&\int d^d x\sqrt{g}[\Tr\left(\pa^{\mu}H^{\dagger}\pa_{\mu}H\right)+u\Tr\left(H^{\dagger}HH^{\dagger}H\right)+v\left[\Tr\left(H^{\dagger}H\right)\right]^2\nn
&+m^2\Tr\left(H^{\dagger}H\right)],
\end{align}
where $g$ denotes the metric determinant and $m^2=\left(\frac{d-2}{2R}\right)^2$ is the conformal coupling required by Weyl invariance. We consider a stationary configuration parametrised by
\be
H_0(\tau)=e^{iM\tau}B
\label{clas1}
\ee
where $M$, $B$ are constant $n\times n$ matrices. 
We parametrise $M$ and $B$ by
\begin{align}
B=&\hbox{diag}\{b,b,0\ldots0\},\nn
M=&-i\hbox{diag}\{\mu,-\mu,0,\ldots0\},
\label{clas2}
\end{align}
and consider an operator of charge $Q$ given by 
\be
T_Q=(H_{11}H^*_{22})^{2J},
\ee
where $Q=4J$. Note that our definition of $M$ in Eq.~\eqref{clas1} differs by a factor of 2 from that in Ref.~\cite{sann2}, resulting in our definitions of $\mu$ also differing by 2.
The parameter $\mu$ may be regarded as a chemical potential, satisfying
\begin{align}
J=&V\mu b^2,\nn
\mu^2=&2(u+2v)b^2+m^2,
\label{Jmu}
\end{align}
where $V$ is the cylinder volume.
In four dimensions this results in the equation for $\mu$
\be
R\mu=\frac{3^{\frac13}+x^{\frac23}}{3^{\frac23}x^{\frac13}},
\label{mudefa}
\ee
where 
\be
x=36\Jcal+\sqrt{1296\Jcal^2-3}
\label{xdef}
\ee
with
\be
\Jcal=2J\frac{\alpha_h+\alpha_y}{N}.
\ee
For simplicity we give in Eq.~\eqref{mudefa} the result for $d=4$. 
An operator with scaling dimension $\Delta$ corresponds to a state with energy $E=\frac{\Delta}{R}$, which may be computed from a fixed-charge path integral on the cylinder. The fixed charge path integral is equivalent to an unconstrained path integral with the addition of boundary terms to the cylinder lagrangian $S_{\rm{cyl}}$, to give an effective action $S_{\rm{eff}}$.
The scaling dimension $\Delta_{T_{Q}}$ of the operator $T_Q$ is expanded in a charge expansion as 
\be
\Delta_{T_{Q}}=Q\left(\frac d2-1\right)+\gamma_{T_{Q}}=\Delta_{LO}+\Delta_{NLO}+\ldots,
\label{Tscal}
\ee
where $\gamma_{T_{Q}}$ is the anomalous dimension of the operator $T_Q$.
The leading contribution $\Delta_{LO}$ may be written as $\Delta_{LO}=RE_{LO}$ in terms of the classical energy $E_{LO}$, obtained by substituting the classical solution Eqs.~\eqref{clas1}, \eqref{clas2} into $S_{\rm{eff}}$, with the boundary terms now resulting in an additional $4\mu^2 b^2$. The leading quantum correction $\Delta_{NLO}=RE_{NLO}$ is obtained from a functional determinant, which may be derived from the  modes associated with the quadratic fluctuations around the classical solution. 
The dispersion relations for the modes $\omega_i(l)$, and their corresponding multiplicities $g_i(N)$, are given by
\begin{align}
\omega_1(l)=&\sqrt{J_l^2+\mu^2},\quad g_1=4N-8 ,\nn
\omega_2(l)=&\sqrt{J_l^2+\mu^2-a\mutilde^2}, \quad g_2=2(N-2)^2,\nn
\omega_{3,4}(l)=&\sqrt{J_l^2+\mu^2}\mp\mu, \quad g_{3,4}=2N-3,\nn
\omega_{5,6}(l)=&\sqrt{J_l^2+\mu^2+2a\mutilde^2}\pm\mu,\quad g_{5,6}=1,\nn
\omega_{7,8}(l)=&\sqrt{J_l^2+3\mu^2-m^2\pm\sqrt{4J_l^2\mu^2+(3\mu^2-m^2)^2}},\quad g_{7,8}=1,\nn
\omega_{9,10}(l)=&\left(J_l^2+2\mu^2+a\mutilde^2\pm\sqrt{4\mu^2J_l^2+(2\mu^2+a\mutilde^2)^2}\right)^{\frac12},\quad g_{9,10}=1,
\label{omdef}
\end{align}
where
\be
a=\frac{u}{u+2v}=\frac{\alpha_h}{\alpha_h+\alpha_y},
\label{adef}
\ee
and where 
\be
\mutilde^2=\mu^2-m^2
\label{mutdef}
\ee
and 
\be
J_l^2=\frac{l(l+d-2)}{R^2}
\label{Jdef}
\ee
is the eigenvalue of the Laplacian on the sphere. 
Then using Eq.~\eqref{Jmu}, the result may be written in the form
\be
\Delta_{LO}=\frac{Q}{4}\left[3R\mu +\frac{1}{R\mu}\right],
\label{DelmQ}
\ee
where $R\mu$ is given by Eq.~\eqref{mudefa}. Its expansion for small $\alpha_hQ$, $\alpha_yQ$ takes the form
\begin{align}
\Delta_{LO}=&Q\Bigl[1+\frac{Q(\alpha_h+\alpha_y)}{N}-2\frac{Q^2(\alpha_h+\alpha_y)^2}{N^2}+8\frac{Q^3(\alpha_h+\alpha_y)^3}{N^3}\nn
&-42\frac{Q^4(\alpha_h+\alpha_y)^4}{N^4}+\ldots\Bigr],
\label{lead}
\end{align}
where we have displayed the expansion up to fourth order, corresponding to four loops in perturbation theory.

As mentioned earlier, the mapping to the cylinder (along with other technical simplifications\cite{Bad2}) relies on conformal invariance. At leading order, the Lagrangian Eq.~\eqref{lag4} is Weyl invariant for any values of the couplings, and so Eq.~\eqref{DelmQ} is valid for arbitrary couplings; but now to  proceed to next-to-leading order, we must assume that we are at the conformal fixed point $\alpha_h=\alpha_h^*$, $\alpha_y=\alpha_y^*$. 
We then find that $\Delta_{NLO}$ is given by
\be
\Delta_{NLO}=\frac{1}{2}\sum_{l=0}^{\infty}\sigma_l
\label{lsum}
\ee
where
\begin{align}
\sigma_l=Rn_l\sum_{i=1}^{10}g_i(N)\omega_i(l).
\label{ONDel}
\end{align}
 Here
\be
n_l=\frac{(2l+d-2)\Gamma(l+d-2)}{\Gamma(l+1)\Gamma(d-1)}
\label{nldef}
\ee
 is the multiplicity of the Laplacian on the $d$-dimensional sphere, and $\omega_i(l)$, $g_i(N)$ are the modes and their multiplicities defined as in Eqs.~\eqref{omdef}, but now evaluated at the fixed point.
For the small $\alpha_h^*Q$, $\alpha_y^*Q$ computation, we need to isolate the divergent contribution in the sum in Eq.~\eqref{lsum}. We use the large-$l$ expansion of $\sigma_l$,
\be
\sigma_l=\sum_{n=1}^{\infty}c_nl^{d-n}
\label{largel}
\ee
with
\begin{align}
c_1=&2N^2,\quad c_2=6N^2,\nn
c_3=&\frac{1}{\alpha_h^*+\alpha_y^*}[(R\mu^*)^2(N^2\alpha_y^*+4N\alpha_h^*+\alpha_y^*)+N^2(6\alpha_h^*+5\alpha_y^*)-4N\alpha_h^*-\alpha_y^*],\nn
c_4=&\frac{1}{\alpha_h^*+\alpha_y^*}[(R\mu^*)^2(N^2\alpha_y^*+4N\alpha_h^*+\alpha_y^*)+N^2(2\alpha_h^*+\alpha_y^*)-4N\alpha_h^*-\alpha_y^*],\nn
c_5=&-\frac{[(R\mu^*)^2-1]^2N}{8(\alpha_h^*+\alpha_y^*)^2}G(\alpha^*_h,\alpha^*_y)(1-\gamma\epsilon)\nn
&-\frac{\epsilon}{24(\alpha_h^*+\alpha_y^*)^2}\Bigl\{
N^2[9(R\mu^*)^4\alpha_y^{*2}+2(R\mu^*)^2\alpha_y^*(5\alpha_h^*+2\alpha_y^*)\nn
&+\frac{1}{5}(62\alpha_h^{*2}+74\alpha_h^*\alpha_y^*-3\alpha_y^{*2})]\nn
&+4[(R\mu^*)^2-1]\alpha_h^*N[9(R\mu^*)^2(\alpha_h^*+2\alpha_y^*)+13\alpha_h^*+16\alpha_y^*]\nn
&+2[(R\mu^*)^2-1][9(R\mu^*)^2(6\alpha_h^{*2}+3\alpha_h^*\alpha_y^*+2\alpha_y^{*2})+18\alpha_h^{*2}+14\alpha_h^*\alpha_y^*+11\alpha_y^{*2}]
\Bigr\}\nn
&+\Ocal(\epsilon^2),
\label{cvals}
\end{align}
where for future convenience we define
\be
G(\alpha_h,\alpha_y)=\frac2N[(N^2+4)\alpha_y^{2}+2(4N+3)\alpha_y\alpha_h+4(N+3)\alpha_h^{2}].
\label{Gdef}
\ee
The $\Ocal(\epsilon^2)$ in $c_5$ refers to powers of $\epsilon$ appearing explicitly; neglecting the fact that the fixed point values $\alpha_h^*$, $\alpha_y^*$ are themselves $\Ocal(\epsilon)$. We also compute $\mu^*$ using Eq.~\eqref{mudefa} but similarly evaluated at the conformal fixed point.
We can write Eq.~\eqref{lsum} in the form 
\be
\Delta_{NLO}=\rho+\frac{1}{2}\sum_{l=1}^{\infty}\sigmabar_l.
\label{delrho}
\ee
Here
\be
\rho=\frac12\sum_{n=1}^5c_n\sum_{l=1}^{\infty}l^{d-n}+\frac12\sigma_0
\label{rhodef}
\ee
and
\be
\sigmabar_l=\sigma_l-\sum_{n=1}^5c_nl^{d-n}
\label{sigbar}
\ee
We have now isolated the divergent contributions in $\rho$ in order to evaluate them separately, as explained in Refs.~\cite{Bad2} and \cite{sann}. The sum over $\sigmabar_l$ is convergent and may be evaluated directly in $d=4$. The terms with $c_{1-5}$ in Eq.~\eqref{rhodef} are evaluated using
\be
\sum_{l=1}^{\infty}l^s=\zeta(-s)
\label{zdef}
\ee
where $\zeta(x)$ is the Riemann $\zeta$-function. The terms in Eq.~\eqref{rhodef} with $n=1\ldots4$ may be evaluated setting $d=4$ and using 
\be
\zeta(-s)=(-1)^s\frac{B_{s+1}}{s+1},\quad s=0,1,2\ldots ;
\label{bern}
\ee
while for the term with $n=5$ we use
\be
\zeta(1+\epsilon)=\frac{1}{\epsilon}+\gamma.
\label{zpole}
\ee
The pole here in $\epsilon$ cancels against the poles arising from the bare couplings $\alpha_h^{B(1)}$, $\alpha_y^{B(1)}$ in the leading-order energy $E_{LO}$ (this being the occasion mentioned earlier where we need to reintroduce them). This follows from the fact that $G(\alpha_h,\alpha_y)$ in Eq.~\eqref{Gdef} satisfies
\be
\alpha_h^{B(1)}+\alpha_y^{B(1)}=\alpha_h+\alpha_y+\frac{1}{\epsilon}G(\alpha_h,\alpha_y),
\label{Gprop}
\ee
after substituting the expressions for the bare couplings in Eq.~\eqref{bare}. We then obtain a finite result as $\epsilon\rightarrow0$. The terms in $\gamma$ clearly also cancel in this limit. After some calculation we find 
\begin{align}
\rho=&-\frac{1}{16(\alpha_h^*+\alpha_y^*)^2}\{3[4(N+3)\alpha_h^{*2}+2(4N+3)\alpha_h^*\alpha_y^*+(N^2+4)\alpha_y^{*2}](R\mu^*)^4\nn
&+2[12(N-1)\alpha_h^{*2}+2(2N^2+4N-1)\alpha_h^*\alpha_y^*+3N^2\alpha_y^{*2}](R\mu^*)^2\nn
&+[4(4N^2-9N-3)\alpha_h^{*2}+2(12N^2-20N-7)\alpha_h^*\alpha_y^*+(7N^2-12)\alpha_y^{*2}]\}\nn
&+\frac12\sigma_0,
\end{align}
with $\omega_i(l)$ given by Eq.~\eqref{omdef}.
Performing the sum over $\sigmabar_l$, putting everything together, and expanding in $\alpha_h^*Q$, $\alpha_y^*Q$ up to fourth order (again corresponding to four loops in perturbation theory), we obtain \cite{sann2}
\begin{align}
\Delta_{NLO}=&
-\frac{Q}{N}\frac{2(2N+7)\alpha_h^{*2}+(8N+9)\alpha_h^*\alpha_y^*+(N^2+5)\alpha_y^{*2}}{\alpha_y^*+\alpha_h^*}\nn
&-\frac{Q^2}{N^2}\left[2\left(2N-1)\right)\alpha_h^{*2}+4\left(2N-3\right)\alpha_h^*\alpha_y^*+\left(N^2-3\right)\alpha_y^{*2}\right]\nn
&+\frac{Q^3}{N^3}\Bigl\{4\left[4(N+9)\alpha_h^{*3}+6(2N+7)\alpha_y^*\alpha_h^{*2}+6(2N+3)\alpha_h^*\alpha_y^{*2}+(N^2+7)\alpha_y^{*3}\right]\zeta_3\nn
&+2\left[4(N-8)\alpha_h^{*2}+2(4N-25)\alpha_y^*\alpha_h^*+(N^2-18)\alpha_y^{*2}\right](\alpha_h^*+\alpha_y^*)\Bigr\}\nn
&-\frac{Q^4}{N^4}\Bigl\{\Bigl[4(24N+197)\alpha_h^{*4}+8(48N+209)\alpha_y^*\alpha_h^{*3}+48(12N+25)\alpha_y^{*2}\alpha_h^{*2}\nn
&+8(3N^2+36N+56)\alpha_y^{*3}\alpha_h^*+2(12N^2+65)\alpha_y^{*4}\Bigr]\zeta_3\nn
&+10\Bigl[2(2N+53)\alpha_h^{*4}+16
(N+11)\alpha_y^*\alpha_h^{*3}
+12(2N+11)\alpha_y^{*2}\alpha_h^{*2}\nn
&+8(2N+7)\alpha_y^{*3}\alpha_h^*
+(N^2+15)\alpha_y^{*4}\Bigr]\zeta_5\nn
&+8(4N-69)\alpha_h^{*4}+8(16N-233)\alpha_y^*\alpha_h^{*3}+2(4N^2+80N-1187)\alpha_y^{*2}\alpha_h^{*2}\nn
&+8(2N^2+8N-169)\alpha_y^{*3}\alpha_h^*+4(2N^2-73)\alpha_y^{*4}\Bigr\}
+\ldots
\label{nlead}
\end{align}
Writing $\Delta_S=\Delta_{LO}+\Delta_{NLO}$ for our full semiclassical result, we find from Eqs.~\eqref{lead}, \eqref{nlead}
\begin{align}
\Delta_S=&Q+\frac{Q}{N}\bigg\{Q(\alpha^*_h+\alpha^*_y)\nn
&-\frac{2(2N+7)\alpha_h^{*2}+(8N+9)\alpha_h^*\alpha_y^*+(N^2+5)\alpha_y^{*2}}{\alpha_y^*+\alpha_h^*}\bigg\}\nn
&-\frac{Q^2}{N^2}\bigg\{2Q(\alpha^*_h+\alpha^*_y)^2\nn
&+\left[2\left(2N-1)\right)\alpha_h^{*2}+4\left(2N-3\right)\alpha_h^*\alpha_y^*+\left(N^2-3\right)\alpha_y^{*2}\right]\bigg\}\nn
&+\frac{Q^3}{N^3}\bigg\{8Q(\alpha^*_h+\alpha^*_y)^3\nn
&+4\left[4(N+9)\alpha_h^{*3}+6(2N+7)\alpha_y^*\alpha_h^{*2}+6(2N+3)\alpha_h^*\alpha_y^{*2}+(N^2+7)\alpha_y^{*3}\right]\zeta_3\nn
&+2\left[4(N-8)\alpha_h^{*2}+2(4N-25)\alpha_y^*\alpha_h^*+(N^2-18)\alpha_y^{*2}\right](\alpha_h^*+\alpha_y^*)\bigg\}\nn
&-\frac{Q^4}{N^4}\bigg\{42Q(\alpha^*_h+\alpha^*_y)^4\nn
&+\Bigl[4(24N+197)\alpha_h^{*4}+8(48N+209)\alpha_y^*\alpha_h^{*3}+48(12N+25)\alpha_y^{*2}\alpha_h^{*2}\nn
&+8(3N^2+36N+56)\alpha_y^{*3}\alpha_h^*+2(12N^2+65)\alpha_y^{*4}\Bigr]\zeta_3\nn
&+10\Bigl[2(2N+53)\alpha_h^{*4}+16
(N+11)\alpha_y^*\alpha_h^{*3}
+12(2N+11)\alpha_y^{*2}\alpha_h^{*2}\nn
&+8(2N+7)\alpha_y^{*3}\alpha_h^*
+(N^2+15)\alpha_y^{*4}\Bigr]\zeta_5\nn
&+8(4N-69)\alpha_h^{*4}+8(16N-233)\alpha_y^*\alpha_h^{*3}+2(4N^2+80N-1187)\alpha_y^{*2}\alpha_h^{*2}\nn
&+8(2N^2+8N-169)\alpha_y^{*3}\alpha_h^*+4(2N^2-73)\alpha_y^{*4}\bigg\}
+\ldots
\label{Delfull}
\end{align}
 Finally, we note that the $U(1)$ results of Ref.~\cite{Bad2} may be recovered by setting $\alpha_h=0$, $N=1$, $Q=n$, and $\alpha_y=\frac{\lambda}{32\pi^2}$.
\section{The diagrammatic calculation}
In this section we carry out the perturbative calculation of the anomalous dimension $\gamma_{T_Q}$ on the left-hand side of Eq.~\eqref{Tscal}, to confirm the result displayed in Eq.~\eqref{Delfull}; namely the semiclassical result at leading and next-to-leading order in $Q$ up to four-loop level.

\begin{figure}
\centering
\includegraphics[width=0.5\columnwidth]{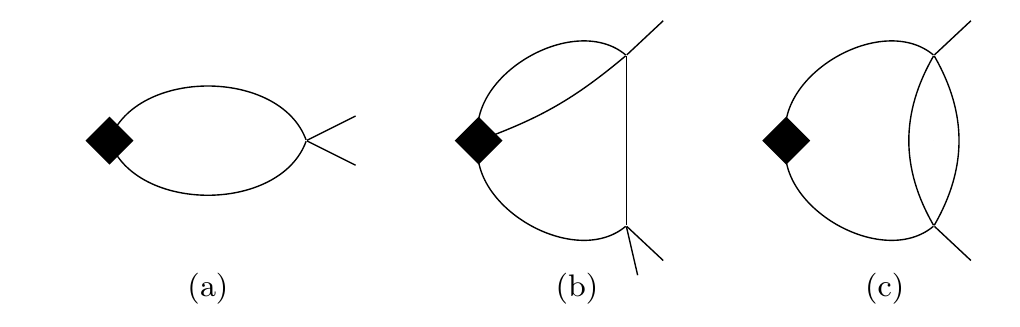}
\caption{One- and two-loop diagrams for $\gamma_{T_{Q}}$}\label{diagtwo}
\end{figure}

The one-loop contribution to $\gamma_{T_{Q}}$ comes solely from the diagram depicted in Fig.~\ref{diagtwo}(a). Strictly the propagators should be directional, with the two ends corresponding to $H$, $H^{\dagger}$ respectively; but we have suppressed the directional arrows for convenience. Each of our graphs in general might thus correspond to several distinct ways of assigning the arrows. Conversely, at three and higher loops one can draw a small number of graphs for which there is no way of assigning the arrows; we have naturally suppressed these. In fact, at least up to four loops the set of contributing graphs turns out to be exactly the same as in the $U(1)$ and $O(N)$ cases. It is not obvious to us why this is so, though possibly it can be traced back to the equivalence mentioned in Ref.~\cite{sann2} between the $U(N)\times U(N)$ results for $\alpha_h=0$, and those for the $O(2N^2)$ theory.

We give the diagrammatic results in the form of a table, where the first column labels the diagram; the second column contains the product of the appropriate symmetry factor and a coupling-dependent term; and the third column contains the simple pole residues for the Feynman diagram. The one, two and three-loop results are all listed in Table~\ref{3loop}. For the one-loop diagram, the coupling-dependent term is given by
\be
A_1=\frac{2Q^2(\alpha_h+\alpha_y)}{N}-\frac{2Q(2\alpha_h+\alpha_y)}{N},
\label{gamone}
\ee
and the one-loop contribution to $\gamma^{(1)}_{T_Q}$ is therefore given by
\be
\gamma^{(1)}_{T_Q}=\frac{Q^2(\alpha_h+\alpha_y)}{N}-\frac{Q(2\alpha_h+\alpha_y)}{N}.
\label{gamone}
\ee
As mentioned before, the derivation of the semiclassical result relied on working at the conformal fixed point. However, surprisingly, at two, three and four loops we will see that the functional forms of the semiclassical and perturbative results agree for general $\alpha_h$, $\alpha_y$ and not just on substitution  of $\alpha_h=\alpha_h^*$, $\alpha_y=\alpha_y^*$, with  $\alpha_h^*$, $\alpha_y^*$ defined by Eq.~\eqref{gfix}. It is only at one loop where the agreement only holds at the fixed point. Specifically, the $\Ocal(Q\alpha_h^0\epsilon)$, $\Ocal(Q\alpha_y^0\epsilon)$ and  $\Ocal(Q\alpha_h)$, $\Ocal(Q\alpha_y)$ terms appearing at the classical and one-loop perturbative levels
in $Q\left(\frac d2-1\right)+\gamma^{(1)}_{T_{Q}}$ on the left-hand side of Eq.~\eqref{Tscal} (as given in Eq.~\eqref{gamone}) only agree with those in $\Delta_{LO}+\Delta_{NLO}$ on the right-hand side of Eq.~\eqref{Tscal} (as obtained from Eq.~\eqref{Delfull}) after substituting the fixed point values satisfying Eq.~\eqref{gfix}. It is easy to see from Eq.~\eqref{betas} that the fixed point values satisfy
\be
\epsilon(\alpha_y^*+\alpha_h^*)=G(\alpha^*_h,\alpha^*_y),
\label{grel}
\ee
where $G(\alpha_h,\alpha_y)$ is defined in Eq.~\eqref{Gdef} (this identity is of course related to Eq.~\eqref{Gprop}).
Using this identity, it is straightforward to check that the two sides of Eq.~\eqref{Tscal} agree up to the one-loop level\cite{sann2}.
 In this case, specialising to the fixed point has induced a mixing between the classical and one-loop $\Ocal(Q)$ terms. However this mixing only occurs between the classical and one-loop results, due to the explicit $\Ocal(\epsilon)$ terms on the left-hand side of Eq.~\eqref{Tscal}.

The leading $\Ocal(Q^3)$ two-loop contribution to $\gamma_{T_{Q}}$ comes purely from the diagram depicted in Fig.~\ref{diagtwo}(b) (with three lines emerging from the $T_{Q}$ vertex), while the next-to-leading $\Ocal(Q^2)$ contributions are generated by this diagram together with those in Fig.~\ref{diagtwo}(c) (with two lines emerging from the $T_{Q}$ vertex). The two-loop coupling-dependent terms in Table~\ref{3loop} are given by
\begin{align}
B_1=&4(\alpha_h+\alpha_y)^2\frac{Q^3}{N^2}-4(3\alpha_y^2+8\alpha_y\alpha_h+6\alpha_h^2)\frac{Q^2}{N^2}+\ldots,\\
B_2=&2[2(2N+5)\alpha_h^2+4(2N+1)\alpha_y\alpha_h+(N^2+3)\alpha_y^2]\frac{Q^2}{N^2}+\ldots,
\label{bdefs}
\end{align}
where the ellipses in the coupling-dependent terms here and later represent terms of lower order in $Q$.
Consequently the leading and next-to-leading contributions to $\gamma^{(2)}_{T_Q}$ are given by
\begin{align}
\gamma^{(2)}_{T_Q}=&-\frac{2Q^3(\alpha_h+\alpha_y)^2}{N^2}\nn
&-\frac{Q^2}{N^2}\left[2\left(2N-1\right)\alpha_h^2+4\left(2N-3\right)\alpha_h\alpha_y+\left(N^2-3\right)\alpha_y^2\right]+\ldots
\end{align}
 in accord with the semiclassical results in Eq.~\eqref{Delfull}. As emphasised earlier, this agreement holds for general $\alpha_h$, $\alpha_y$ and not just at the conformal fixed point. This is because at two and higher loops, in contrast to what we saw at one loop, specialising to the fixed point values $\alpha_h=\alpha_h^*$, $\alpha_y=\alpha_y^*$ satisfying Eq.~\eqref{gfix} does not induce any mixing between leading or next-to-leading terms at different loop orders.  Therefore if Eq.~\eqref{Tscal} holds at the fixed point, it must also hold in general. This point was already made in Ref.~\cite{sann2}, though there the argument was used to predict the full (i.e. beyond next-to-leading order) two-loop anomalous dimension - see the Conclusions for further discussion of this.

\begin{figure}
\centering
\includegraphics[width=0.75\columnwidth]{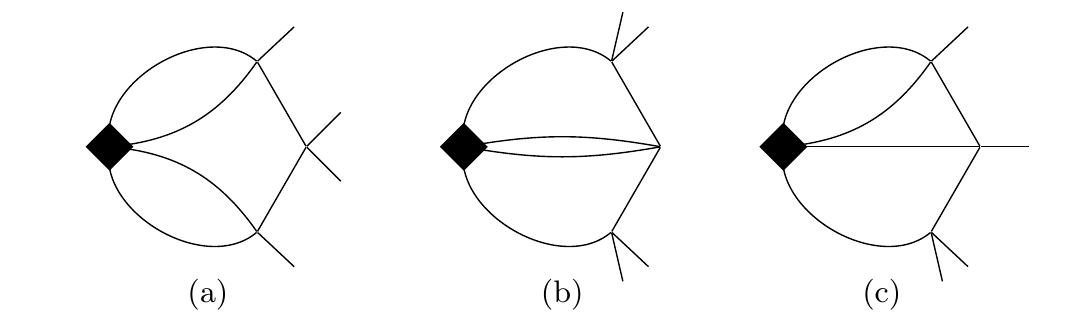}
\caption{Three-loop diagrams for $\gamma_{T_{Q}}$ contributing at leading $n$}\label{diagthree}
\end{figure}
\begin{figure}
\centering
\includegraphics[width=0.75\columnwidth]{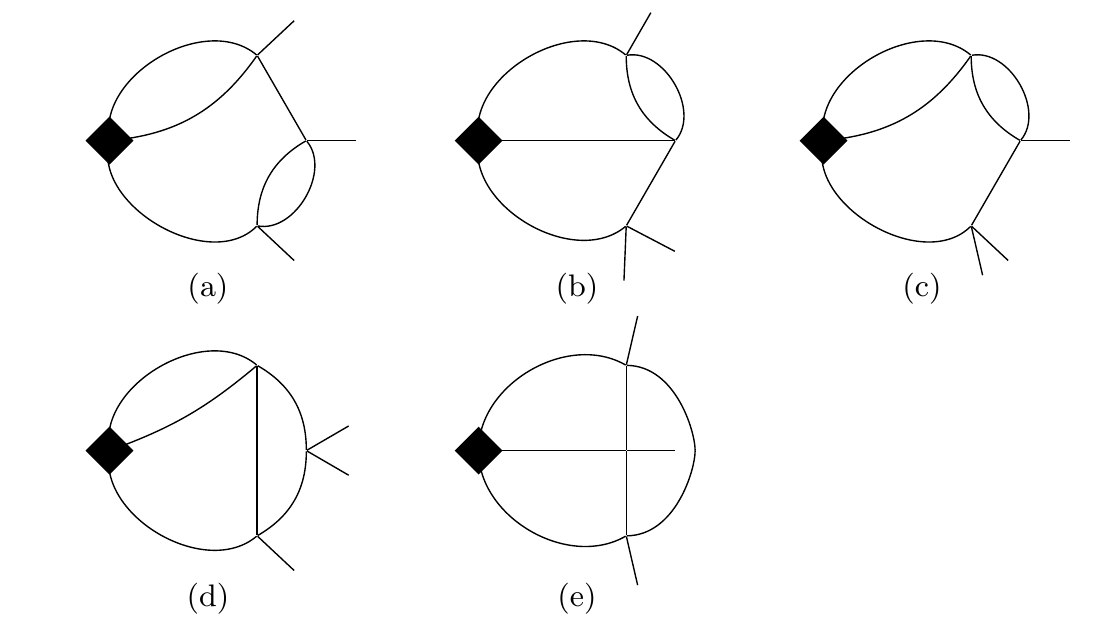}
\caption{Three-loop diagrams for $\gamma_{T_{Q}}$ contributing at next-to-leading $n$}\label{diagthreea}
\end{figure}

The leading $\Ocal(Q^4)$ three-loop contributions to $\gamma_{T_{Q}}$ come purely from the diagrams depicted in Fig.~\ref{diagthree} (with four lines emerging from the $T_{Q}$ vertex), while the next-to-leading $\Ocal(Q^3)$ contributions are generated by these diagrams together with those in Fig.~\ref{diagthreea} (with three lines emerging from the $T_{Q}$ vertex).
\begin{table}
\begin{center}
\begin{tabular}{|c|c|c|} \hline
&&\\
Graph & Symmetry and &Simple Pole\\
&Coupling Factors&\\
&&\\
\hline
&&\\
1(a)&$\frac{1}{4}A_1$&$2$\\
&&\\
\hline
&&\\
1(b)&$-\frac{1}{4}B_1$&$\frac12$\\
&&\\
\hline
&&\\
1(c)&$-\frac{1}{4}B_2$&$\frac12$\\
&&\\
\hline
&&\\
2(a)&$\frac{1}{16}C_1$&$-\frac23$\\
&&\\
\hline
&&\\
2(b)&$\frac{1}{16}C_1$&$\frac23$\\
&&\\
\hline
&&\\
2(c)&$\frac{1}{4}C'_1$&$\frac43$\\
&&\\
\hline
&&\\
3(a)&$\frac{1}{4}C_2$&$-\frac23$\\
&&\\
\hline
&&\\
3(b)&$\frac{1}{4}C_2$&$\frac43$\\
&&\\
\hline
&&\\
3(c)&$\frac{1}{8}C_3$&$-\frac23$\\
&&\\
\hline
&&\\
3(d)&$\frac{1}{4}C'_3$&$\frac43$\\
&&\\
\hline
&&\\
3(e)&$\frac{1}{6}C_4$&$4\zeta_3$\\
&&\\
\hline
\end{tabular}
\caption{\label{3loop}Three-loop results from Figs.~\ref{diagthree}, \ref{diagthreea}}
\end{center}
\end{table}
The simple pole contributions from individual three-loop diagrams may be extracted from Ref.~\cite{kaz} and are listed in Table~\ref{3loop}, together with the corresponding symmetry factors, and a coupling-dependent term $C_{1-4}$. The latter are given by
\begin{align}
C_1=&8\frac{Q^4}{N^3}(\alpha_h+\alpha_y)^3-8\frac{Q^3}{N^3}(3\alpha_h+2\alpha_y)(4\alpha_h^2+6\alpha_h\alpha_y+3\alpha_y^2)+\ldots,\nn
C_1'=&C_1-4\frac{Q^3}{N^3}\alpha_h^2\alpha_y+\ldots,\nn
C_2=&2\frac{Q}{N}(\alpha_h+\alpha_y)B_2+\ldots,\nn
C_3=&8\frac{Q^3}{N^3}(\alpha_h+\alpha_y)(2\alpha_h^2+2\alpha_h\alpha_y+\alpha_y^2)+\ldots,\nn
C'_3=&C_2-4\frac{Q^3}{N^3}\alpha_h^2\alpha_y+\ldots,\nn
C_4=&2\frac{Q^3}{N^3}[N^2\alpha_y^3+4N\alpha_h(\alpha_h^2+3\alpha_h\alpha_y+3\alpha_y^2)\nn
&+36\alpha_h^3+42\alpha_h^2\alpha_y+18\alpha_h\alpha_y^2+7\alpha_y^3]+\ldots,
\label{cdefs}
\end{align}
where $B_2$ is defined in Eq.~\eqref{bdefs}.
 We shall see shortly that relations like that between $C_2$ and $B_2$ occur more frequently at the next loop order. A necessary but not sufficient condition for a relation of this kind seems to be that a diagram containing the lower-loop quantity (Fig.~1(c) in this case, as we see from Table~\ref{3loop}) may be formed by removing one of the lines emerging from the $T_Q$ vertex (the ``lozenge'') in a diagram containing the higher-loop quantity (Fig.~3(a) or 3(b) in this case). The line removed may incorporate a four-point vertex with two external lines, in which case an extra external line is added at the end of the removed line; otherwise, when the line is removed, the external line at its end is also removed. We emphasise that this relation only holds true at leading order in $Q$; and indeed a similar but somewhat trivial relation holds between the leading terms in $C_1$, $C_1'$ and $B_1$ (defined in Eq.~\eqref{bdefs}), and also between $B_1$ and $A_1$ in Eq.~\eqref{gamone}.
When the three-loop contributions from the table are added and multiplied by a loop factor of 3, the leading and non-leading three-loop contributions to  $\gamma_{T_Q}$ are found to be 
\begin{align}
\gamma^{(3)}_{T_Q}=&8\frac{Q^4(\alpha_h+\alpha_y)^3}{N^3}\nn
&+\frac{Q^3}{N^3}\Bigl\{4\left[4(N+9)\alpha_h^3+6(2N+7)\alpha_y\alpha_h^2+6(2N+3)\alpha_h\alpha_y^2+(N^2+7)\alpha_y^3\right]\zeta_3\nn
&+2\left[4(N-8)\alpha_h^2+2(4N-25)\alpha_y\alpha_h+(N^2-18)\alpha_y^2\right](\alpha_h+\alpha_y)\Bigr\}+\ldots
\end{align}
once again in accord with the semiclassical results in Eqs.~\eqref{Delfull}, for general $\alpha_h$, $\alpha_y$.

\begin{figure}
\centering
\includegraphics[width=\columnwidth]{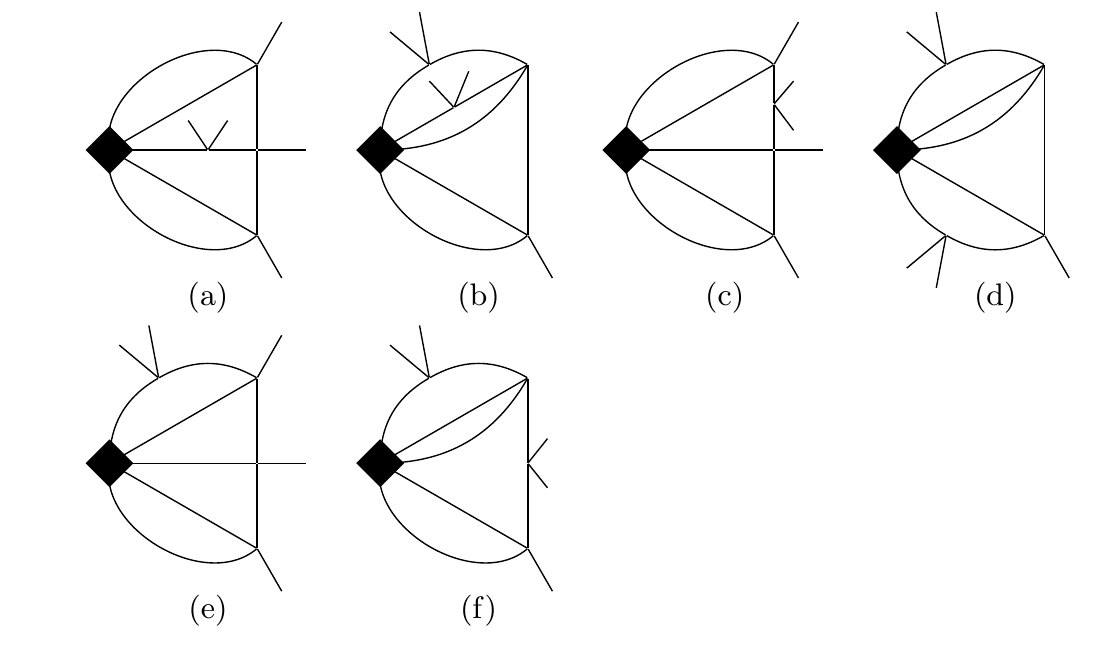}
\caption{Four-loop diagrams for $\gamma_{T_{Q}}$ contributing at leading $n$}\label{diagfour}
\end{figure}

\begin{figure}
\centering
\includegraphics[width=\columnwidth]{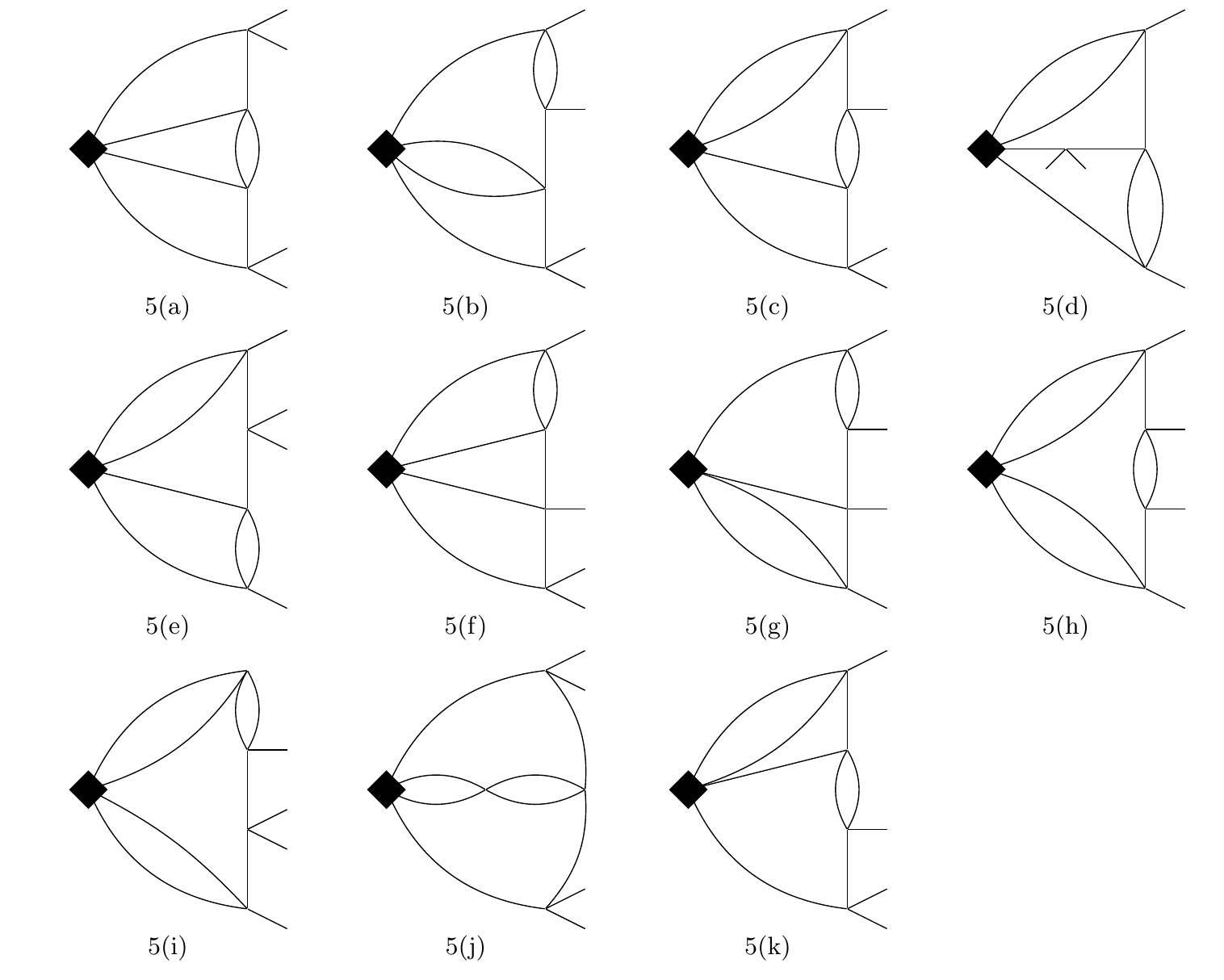}
\caption{Four-loop diagrams for $\gamma_{T_{Q}}$ contributing at next-to-leading $n$}\label{diagfoura}
\end{figure}

The leading $\Ocal(Q^5)$ four-loop contributions to $\gamma_{T_{Q}}$ come purely from the diagrams depicted in Fig.~\ref{diagfour} (with five lines emerging from the $T_{Q}$ vertex), while the next-to-leading $\Ocal(Q^4)$ contributions are generated by these diagrams together with those in Fig.~\ref{diagfoura} (with four lines emerging from the $T_{Q}$ vertex).
\begin{table}
\begin{center}
\begin{tabular}{|c|c|c|} \hline
&&\\
Graph & Symmetry and&Simple Pole\\
&Coupling Factors&\\
&&\\
\hline
&&\\
4(a)&$-\frac{1}{16}D_1$&$-\frac56$\\
&&\\
\hline
&&\\
4(b)&$-\frac{1}{16}D_1$&$\frac{11}{6}$\\
&&\\
\hline
&&\\
4(c)&$-\frac{1}{8}D_2$&$-\frac23$\\
&&\\
\hline
&&\\
4(d)&$-\frac{1}{8}D_2$&$\frac23$\\
&&\\
\hline
&&\\
4(e)&$-\frac{1}{4}D_3$&$\frac52$\\
&&\\
\hline
&&\\
4(f)&$-\frac{1}{16}D_4$&$-\frac12$\\
&&\\
\hline
\end{tabular}
\caption{\label{4loop}Four-loop results from Fig.~\ref{diagfour}}
\end{center}
\end{table}
\begin{figure}
\centering
\includegraphics[width=\columnwidth]{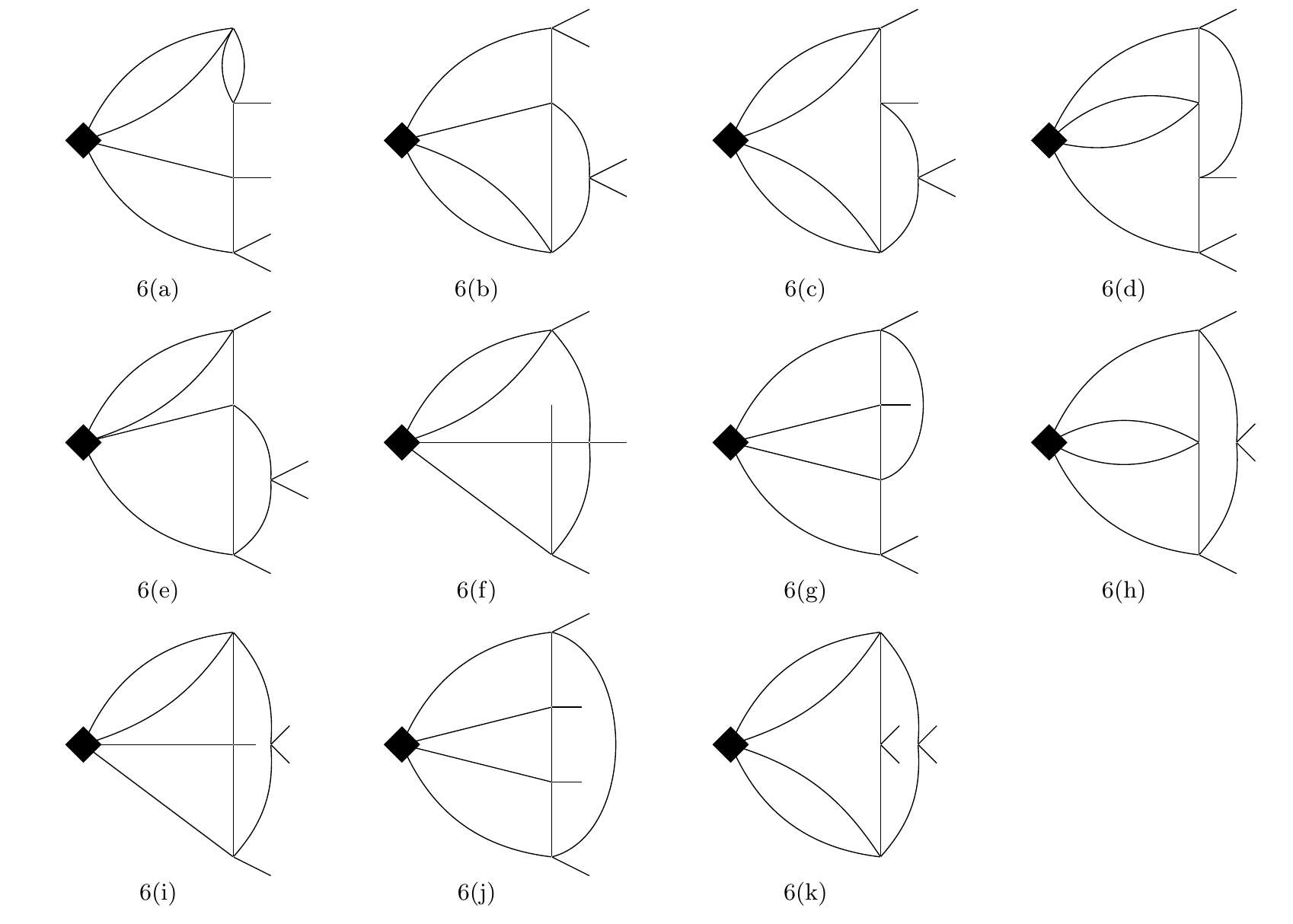}
\caption{Four-loop diagrams for $\gamma_{T_{Q}}$ contributing at next-to-leading $n$ (continued)}\label{diagfourb}
\end{figure}
\begin{table}
\begin{center}
\begin{tabular}{|c|c|c|} \hline
&&\\
Graph & Symmetry and&Simple Pole\\
&Coupling Factors&\\
&&\\
\hline
&&\\
5(a)&$-\frac{1}{16}E_1$&$\frac{1}{6}(11-6\zeta_3)$\\
&&\\
\hline
&&\\
5(b)&$-\frac{1}{8}E_1$&$-\frac{1}{2}$\\
&&\\
\hline
&&\\
5(c)&$-\frac{1}{8}E_1$&$-\frac{1}{6}(5-6\zeta_3)$\\
&&\\
\hline
&&\\
5(d)&$-\frac{1}{8}E_1$&$-\frac{5}{6}$\\
&&\\
\hline
&&\\
5(e)&$-\frac{1}{8}E_1$&$-\frac23$\\
&&\\
\hline
&&\\
5(f)&$-\frac{1}{4}E_1$&$\frac52$\\
&&\\
\hline
&&\\
5(g)&$-\frac{1}{4}E_1$&$-\frac23$\\
&&\\
\hline
&&\\
5(h)&$-\frac{1}{16}E_1$&$\frac12(1-2\zeta_3)$\\
&&\\
\hline
&&\\
5(i)&$-\frac{1}{16}E_2$&$\frac12(1-2\zeta_3)$\\
&&\\
\hline
&&\\
5(j)&$-\frac{1}{32}E_2$&$-2(1-\zeta_3)$\\
&&\\
\hline
&&\\
5(k)&$-\frac{1}{8}E_2$&$-\frac23$\\
&&\\
\hline

\end{tabular}
\caption{\label{4loopa}Four-loop results from Fig.~\ref{diagfoura}}
\end{center}
\end{table}
\begin{table}
\begin{center}
\begin{tabular}{|c|c|c|} \hline
&&\\
Graph & Symmetry and&Simple Pole\\
&Coupling Factors&\\
&&\\
\hline
&&\\
6(a)&$-\frac{1}{8}E_2$&$-\frac{1}{6}(5-6\zeta_3)$\\
&&\\
\hline
&&\\
6(b)&$-\frac{1}{8}E_3$&$\frac{1}{6}(11-6\zeta_3)$\\
&&\\
\hline
&&\\
6(c)&$-\frac{1}{8}E_3$&$-\frac{1}{6}(5-6\zeta_3)$\\
&&\\
\hline
&&\\
6(d)&$-\frac{1}{4}E_3$&$\frac{1}{2}(5-4\zeta_3)$\\
&&\\
\hline
&&\\
6(e)&$-\frac{1}{4}E_3$&$\frac52$\\
&&\\
\hline
&&\\
6(f)&$-\frac{1}{4}E_4$&$\frac32(2\zeta_3-\zeta_4)$\\
&&\\
\hline
&&\\
6(g)&$-\frac{1}{4}E_4$&$\frac32(2\zeta_3+\zeta_4)$\\
&&\\
\hline
&&\\
6(h)&$-\frac{1}{8}E_5$&$-(1-2\zeta_3)$\\
&&\\
\hline
&&\\
6(i)&$-\frac{1}{4}E_5$&$\frac{1}{2}(5-4\zeta_3)$\\
&&\\
\hline
&&\\
6(j)&$-\frac{1}{8}E_6$&$10\zeta_5$\\
&&\\
\hline
&&\\
6(k)&$-\frac{1}{64}E_7$&$-2(1-\zeta_3)$\\
&&\\
\hline
\end{tabular}
\caption{\label{4loopb}Four-loop results from Fig.~\ref{diagfourb}}
\end{center}
\end{table}
The simple pole contributions from the four-loop diagrams in Fig.~\ref{diagfour} were readily evaluated using standard techniques (see for instance Ref.~\cite{klein}). Those from Figs.~\ref{diagfoura}, \ref{diagfourb} may be extracted from Ref.~\cite{kaz}. The contributions from the four-loop diagrams in  Figs.~\ref{diagfour}, \ref{diagfoura} and \ref{diagfourb} are listed in Tables~\ref{4loop}, \ref{4loopa} and \ref{4loopb} respectively, together with the corresponding symmetry factor and coupling-dependent factor $D_{1-4}$ or $E_{1-9}$. The latter are listed below: 
\begin{align}
D_1=&16\frac{Q^5}{N^4}(\alpha_h+\alpha_y)^4-32\frac{Q^4}{N^4}(\alpha_h+\alpha_y)(10\alpha_h^3+21\alpha_h^2\alpha_y+17\alpha_h\alpha_y^2+5\alpha_y^3)+\ldots,\nn
D_2=&D_1+8\frac{Q^4}{N^4}\alpha_h^2\alpha_y(2\alpha_h+\alpha_y)+\ldots,\nn
D_3=&D_1-4\frac{Q^4}{N^4}\alpha_h^2\alpha_y^2+\ldots,\nn
D_4=&D_1+16\frac{Q^4}{N^4}\alpha_h^2\alpha_y(2\alpha_h+\alpha_y)+\ldots,
\label{ddefs}
\end{align}
 and furthermore
\begin{align}
E_1=&2\frac{Q}{N}(\alpha_h+\alpha_y)C_2+\ldots,\nn
E_2=&2\frac{Q}{N}(\alpha_h+\alpha_y)C_3+\ldots,\nn
E_3=&2\frac{Q}{N}(\alpha_h+\alpha_y)C'_3+\ldots,\nn
E_4=&2\frac{Q}{N}(\alpha_h+\alpha_y)C_4+\ldots,\nn
E_5=&4\frac{Q^4}{N^4}(8\alpha_h^4+20\alpha_y\alpha_h^3+25\alpha_y^2\alpha_h^2+16\alpha_y^3\alpha_h+4\alpha_y^4)+\ldots,\nn
E_6=&2\frac{Q^4}{N^4}[2(2N+53)\alpha_h^4+16\alpha_y(N+11)\alpha_h^3+12(2N+11)\alpha_y^2\alpha_h^2\nn
&+8(2N+7)\alpha_y^3\alpha_h+\alpha_y^4(N^2+15)]+\ldots,\nn
E_7=&16\frac{Q^4}{N^4}(2\alpha_h^4+4\alpha_y\alpha_h^3+6\alpha_y^2\alpha_h^2+4\alpha_y^3\alpha_h+\alpha_y^4)+\ldots,\nn
\end{align}
where $C_2$, $C_3$, $C'_3$ and $C_4$ are as defined in Eq.~\eqref{cdefs}.
As anticipated earlier, we see more examples of relations between the coupling-dependent quantities at different loop orders. It is easy to check that the corresponding diagrams follow the rule of thumb described previously. Furthermore, similar trivial relations hold between the leading terms in $D_{1-4}$ in Eq.~\eqref{ddefs} and those in $C_1$, $C'_1$ in Eq.~\eqref{cdefs}.
 When the four-loop results in the tables are added and multiplied by a loop factor of 4, the leading and non-leading four-loop contributions to  $\gamma_{T_{Q}}$ are found to be 
\begin{align}
\gamma^{(4)}_{T_Q}=&-42\frac{Q^5(\alpha_h+\alpha_y)^4}{N^4}\nn
&-\frac{Q^4}{N^4}\Bigl\{\Bigl[4(24N+197)\alpha_h^4+8(48N+209)\alpha_y\alpha_h^3+48(12N+25)\alpha_y^2\alpha_h^2\nn
&+8(3N^2+36N+56)\alpha_y^3\alpha_h+2(12N^2+65)\alpha_y^4\Bigr]\zeta_3\nn
&+10\Bigl[2(2N+53)\alpha_h^4+16
(N+11)\alpha_y\alpha_h^3
+12(2N+11)\alpha_y^2\alpha_h^2\nn
&+8(2N+7)\alpha_y^3\alpha_h
+\alpha_y^4(N^2+15)\Bigr]\zeta_5\nn
&+8(4N-69)\alpha_h^4+8(16N-233)\alpha_y\alpha_h^3+2(4N^2+80N-1187)\alpha_y^2\alpha_h^2\nn
&+8(2N^2+8N-169)\alpha_y^3\alpha_h+4(2N^2-73)\alpha_y^4\Bigr\}+\ldots
\end{align}
once again in accord with the semiclassical results in Eqs.~\eqref{Delfull}, for general $\alpha_h$, $\alpha_y$.

\section{The $d=3$ calculation}
In this section we shall consider the classically scale-invariant $U(N)\times U(N)$ theory in three dimensions. Once again we shall show how the leading and next-to-leading terms in the charge expansion  may be obtained from a semiclassical computation, and verify by comparison with perturbative results. Much of the formalism is a natural adaptation of the corresponding results in four dimensions.

The Lagrangian is given by
\begin{align}
{\cal L} =&\Tr\left(\pa^{\mu}H^{\dagger}\pa_{\mu}H\right)+u\Tr\left(H^{\dagger}HH^{\dagger}HH^{\dagger}H\right)\nn
&+v\left[\Tr\left(H^{\dagger}H\right)\right]^3+w\Tr\left(H^{\dagger}HH^{\dagger}H\right)\Tr\left(H^{\dagger}H\right).
\label{lag3}
\end{align}
We define for convenience
\be
\alpha_h=\frac{uN}{64\pi^2},\quad \alpha_y=\frac{4vN}{64\pi^2}, \quad \alpha_w=\frac{2wN}{64\pi^2}.
\ee
Note that $\alpha_h$ and $\alpha_y$ are defined differently from the case of four dimensions. Also, in $d=3$ the $\beta$-function starts at two-loop order and the model is conformally invariant up to $\Ocal(\alpha_{h,y,w})$ so that at leading and next-to-leading order our results are valid away from the conformal fixed point.
As in the $d=4$ case, we map to a cylinder and consider a stationary solution given by Eqs.~\eqref{clas1}, \eqref{clas2}.
The relation between $J$ and $\mu$ is as in Eq.~\eqref{Jmu} but $\mu$ is now given by
\be
\mu^2=3(u+4v+2w)b^4+m^2,
\label{mu3}
\ee
with of course now a different value for $m^2=\left(\frac{d-2}{2R}\right)^2$ with $d=3$.
Consequently we now find
\be
R\mu=\frac{1}{2\sqrt2}\sqrt{1+\sqrt{1+\Jcal}}
\label{mudefb}
\ee
where
\be
\Jcal=48Q^2\frac{\alpha_h+\alpha_y+\alpha_w}{N}.
\ee
The scaling dimension $\Delta_{T_Q}$ is once again expanded as in Eq.~\eqref{Tscal}, but now we find on computing the classical energy by substituting Eqs.~\eqref{clas1} and \eqref{clas2} into Eq.~\eqref{lag3} and using Eqs.~\eqref{Jmu} and \eqref{mu3} 
\be
\Delta_{LO}=RE_{LO}=\frac{Q}{12}\left[8R\mu +\frac{1}{R\mu}\right],
\label{DelmQ3}
\ee
where $R\mu$ is given by Eq.~\eqref{mudefb}. 
$\Delta_{LO}$ now has an expansion
\be
\Delta_{LO}=Q\left(\frac12+Q^2\frac{\alpha_h+\alpha_w+\alpha_y}{N}-9Q^4\frac{(\alpha_h+\alpha_w+\alpha_y)^2}{N^2}+\ldots\right).
\label{LO3}
\ee
The modes $\omega_i(l)$ and their corresponding multiplicities $g_i(N)$ are given by
\begin{align}
\omega_1(l)=&\sqrt{J_l^2+\mu^2},\quad g_1=4N-8 ,\nn
\omega_2(l)=&\sqrt{J_l^2+\mu^2-\atilde \mutilde^2}, \quad g_2=2(N-2)^2,\nn
\omega_{3,4}(l)=&\sqrt{J_l^2+\mu^2}\mp\mu, \quad g_{3,4}=2N-3,\nn
\omega_{5,6}(l)=&\sqrt{J_l^2+\mu^2+4\abar \mutilde^2}\pm\mu,\quad g_{5,6}=1,\nn
\omega_{7,8}(l)=&\sqrt{J_l^2+4\mu^2-2m^2\pm2\sqrt{J_l^2\mu^2+(2\mu^2-m^2)^2}},\quad g_{7,8}=1,\nn
\omega_{9,10}(l)=&\left(J_l^2+2\mu^2+2\abar\mutilde^2\pm2\sqrt{\mu^2J_l^2+(\mu^2+\abar\mutilde^2)^2}\right)^{\frac12},\quad g_{9,10}=1,
\label{omdefa}
\end{align}
where $\mutilde$ is defined in Eq.~\eqref{mutdef}, and
\begin{align}
\abar=&\frac{u+\frac23w}{u+4v+2w}=\frac{\alpha_h+\frac13\alpha_w}{\alpha_h+\alpha_y+\alpha_w},\nn
\atilde=&\frac{u+\frac43w}{u+4v+2w}=\frac{\alpha_h+\frac23\alpha_w}{\alpha_h+\alpha_y+\alpha_w},
\label{bdefsa}
\end{align}

Notice that the modes in the $d=3$ case in Eq.~\eqref{omdefa} may be simply obtained from those for the $d=4$ case in Eq.~\eqref{omdef}, by replacing $a$ (defined in Eq.~\eqref{adef}) in $\omega_2(l)$ by $\atilde$ (defined in Eq.~\eqref{bdefsa}) and $a$ in $\omega_{5,6}(l)$ and $\omega_{9,10}(l)$ by $2\abar$ (again defined in Eq.~\eqref{bdefsa}).

The next-to-leading contribution to the anomalous dimension is again given by Eqs.~\eqref{lsum}, \eqref{ONDel} with the multiplicity $n_l$ given by Eq.~\eqref{nldef}. 
In the large-$l$ expansion of Eq.~\eqref{largel}, we now have
\begin{align}
c_1=c_2=&4N^2,\nn
c_3=&\frac{1}{6(\alpha_h+\alpha_w+\alpha_y)}\Bigl\{[4(\alpha_w+3\alpha_y)(R\mu)^2
+6\alpha_h+5\alpha_w+3\alpha_y]N^2\nn
&+4(3\alpha_h+2\alpha_w)[4(R\mu)^2-1]N
+2[4(R\mu)^2-1](6\alpha_h+2\alpha_w+3\alpha_y)\Bigr\},\nn
c_4=&\epsilon\left(\frac13N^2-\frac12c_3\right)+\Ocal(\epsilon^2).
\label{cthree}
\end{align}
 We explicitly isolate the divergent contribution to the sum over $l$ in a similar expression to Eq.~\eqref{delrho}:
\be
\Delta_{NLO}=\rho+\frac12\sum_{l=1}^{\infty}\sigmabar_l
\label{delrho3}
\ee
but now valid away from the conformal fixed point and with 
\be
\sigmabar_l=\sigma_l-\sum_{n=1}^4c_nl^{d-n}
\ee
and 
\be
\rho=\frac12\sum_{n=1}^4c_n\sum_{l=1}^{\infty}l^{d-n}+\frac12\sigma_0
\label{rhodef3}
\ee
where the $\omega_i(l)$ are as given in Eq.~\eqref{omdefa}. Here once again the divergent parts have been isolated and the sums over $l$ in Eq.~\eqref{rhodef3} may now be performed. There is a slight subtlety here. With now $d=3-\epsilon$ the sum over $\frac{1}{l^{d-n}}$ for $n=4$ leads using Eqs.~\eqref{zdef}, \eqref{zpole} to a pole in $\epsilon$ which multiplied by $c_4$ in Eq.~\eqref{cthree} leaves a finite remainder. Using also Eq.~\eqref{bern}, we obtain
\be
\rho=-\frac12c_3+\frac12\sigma_0,
\ee
The sum over $\sigmabar_l$ is then finite and we may set $d=3$, whereupon $c_4$ vanishes from $\sigmabar_l$. Expanding in small $\alpha_hQ$, $\alpha_yQ$, $\alpha_wQ$, we find
\begin{align}
\Delta_{NLO}=&-\frac{Q^2}{N}(6\alpha_h+3\alpha_y+4\alpha_w)\nn
&-\frac18\pi^2\frac{Q^4}{N^2}[(3\alpha_y+\alpha_w)^2N^2+4(3\alpha_h+2\alpha_w)(3\alpha_h+6\alpha_y+4\alpha_w)N\nn
&+12(27\alpha_h^2+6\alpha_y^2+8\alpha_w^2+28\alpha_h\alpha_w+15\alpha_h\alpha_y+11\alpha_y\alpha_w)]\nn
&+2\frac{Q^4}{N^2}(72\alpha_h^2+36\alpha_y^2+46\alpha_w^2+114\alpha_h\alpha_w+99\alpha_h\alpha_y+81\alpha_y\alpha_w)+\ldots
\label{NLO3}
\end{align}

\begin{figure}
\centering
\includegraphics[width=0.25\columnwidth]{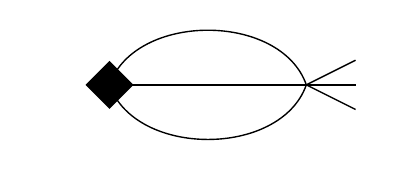}
\caption{Two-loop diagram corresponding to $\gamma_{T_Q}$in $d=3$} \label{diagtwo3}
\end{figure}
\begin{figure}[h]
\centering
\includegraphics[width=0.75\columnwidth]{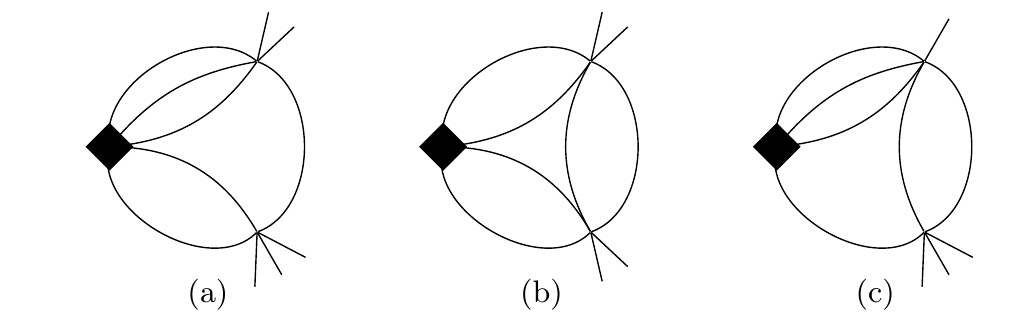}
\caption{Four-loop diagrams corresponding to $\gamma_{T_Q}$ in $d=3$} \label{diagfour3}
\end{figure}

We now perform the perturbative check of Eqs.~\eqref{LO3}, \eqref{NLO3} up to four-loop order (note that in odd dimensions, divergences only appear at even loop order in dimensional regularisation). The two-loop diagram is shown in Fig.~\ref{diagtwo3} and the relevant four-loop diagrams in Fig.~\ref{diagfour3} (Fig.~\ref{diagfour3}(a) providing leading and next-to-leading contributions in $Q$, and Figs.~\ref{diagfour3}(b), (c) giving the remaining next-to-leading order contributions). As before, the lozenge represents the location of the $T_Q$ vertex. 

\begin{table}[h]
\begin{center}
\begin{tabular}{|c|c|c|} \hline
&&\\
Graph & Symmetry and&Simple Pole\\
&Coupling Factors&\\
&&\\
\hline
&&\\
7&$\frac{1}{36}F$&$2$\\
&&\\
\hline
&&\\
8(a)&$-\frac{1}{144}G_1$&$4$\\
&&\\
\hline
&&\\
8(b)&$-\frac{1}{64}G_2$&$\pi^2$\\
&&\\
\hline
&&\\
8(c)&$-\frac{1}{72}G_3$&$4$\\
&&\\
\hline
\end{tabular}
\caption{\label{24loop}Two-loop and four-loop results from Figs.~\ref{diagtwo3}, \ref{diagfour3}}
\end{center}
\end{table}
The coupling-dependent factors in Table~\ref{24loop} are given by
\begin{align}
F=&9\left[\frac{Q^3}{N}(\alpha_h+\alpha_y+\alpha_w)-\frac{Q^2}{N}(6\alpha_h+4\alpha_y+3\alpha_w)\right],\nn
G_1=&9\Bigl[9\frac{Q^5}{N^2}(\alpha_h+\alpha_y+\alpha_w)^2\nn
&-2\frac{Q^4}{N^2}(90\alpha_h^2+45\alpha_y^2+56\alpha_w^2+138\alpha_h\alpha_w+117\alpha_h\alpha_y+99\alpha_y\alpha_w)\Bigr],\nn
G_2=&2\frac{Q^4}{N^2}[(3\alpha_y+\alpha_w)^2N^2+4(3\alpha_h+2\alpha_w)(3\alpha_h+6\alpha_y+4\alpha_w)N\nn
&+6(54\alpha_h^2+12\alpha_y^2+16\alpha_w^2+56\alpha_h\alpha_w+30\alpha_h\alpha_y+22\alpha_y\alpha_w)],\nn
G_3=&9\frac{Q^4}{N^2}(18\alpha_h^2+9\alpha_y^2+10\alpha_w^2+24\alpha_h\alpha_w+18\alpha_h\alpha_y+18\alpha_y\alpha_w).
\end{align}
One might guess at a similar relation between leading-order coupling-dependent factors at successive loops as we found in four dimensions, this time with a relative factor of $9\frac{Q^2}{N}(\alpha_h+\alpha_y+\alpha_w)$; but more evidence would be required to be convincing.
After adding the contributions from the diagrams and multiplying by a loop factor of two or four as appropriate, we find
\begin{align}
\gamma_{T_Q}=&\frac{Q^3}{N}(\alpha_h+\alpha_y+\alpha_w)-Q^2(6\alpha_h+3\alpha_y+4\alpha_w)\nn
&-9\frac{Q^5}{N^2}(\alpha_h+\alpha_y+\alpha_w)^2\nn
&-\frac18\pi^2\frac{Q^4}{N^2}[(3\alpha_y+\alpha_w)^2N^2+4(3\alpha_h+2\alpha_w)(3\alpha_h+6\alpha_y+4\alpha_w)N\nn
&+12(27\alpha_h^2+6\alpha_y^2+8\alpha_w^2+28\alpha_h\alpha_w+15\alpha_h\alpha_y+11\alpha_y\alpha_w)]\nn
&+2\frac{Q^4}{N^2}(72\alpha_h^2+36\alpha_y^2+46\alpha_w^2+114\alpha_h\alpha_w+99\alpha_h\alpha_y+81\alpha_y\alpha_w)+\ldots
\end{align}
We easily see that Eq.~\eqref{Tscal} is satisfied with $\Delta_{LO}$ and $\Delta_{NLO}$ as given by Eqs.~\eqref{LO3}, \eqref{NLO3}. Note that the agreement is for general values of the couplings and not only at the fixed point, even at lowest perturbative order. This is because there is no additional mixing for leading and non-leading $Q$ at the fixed point where $\alpha_{h,y,w}=\Ocal(\epsilon)$; for instance the leading and non-leading terms at lowest order (two loops) are $\Ocal(Q^3)$ and $\Ocal(Q^2)$, whereas the classical terms are $\Ocal(Q)$. We note that the $U(1)$ results of Ref.~\cite{Bad}, \cite{JJ} may be recovered by setting $\alpha_h=\alpha_w=0$, $N=1$, $Q=n$, and $\alpha_y=\frac{\lambda^2}{192\pi^2}$.

\section{Conclusions}

We have extended the investigations of Ref.~\cite{sann2} into large charge operators in scalar $U(N)\times U(N)$ theory. However we have proceeded in a slightly different direction; the authors of Ref.~\cite{sann2} used their leading order (LO) and next-to-leading order (NLO) semiclassical result, combined with a known two-loop result for a particular charge ($Q=2$) to deduce the next-to-next-to-leading order (NNLO) two-loop terms for general $Q$, hence obtaining the full two-loop result for general $Q$. We have taken the approach of using an operator of general charge $Q$ and thereby performing a perturbative check of the LO and NLO semiclassical results up to four-loop order, for general $Q$. This was facilitated by the fact that many of the relevant three-loop and four-loop Feynman diagrams had already been evaluated, in Ref.~\cite{kaz}. This in turn was because the NLO graphs at three loops contained a $T_Q$ vertex (corresponding to the fixed-charge operator) with three internal lines, and the LO graphs at three loops and NLO graphs at four loops contained a $T_Q$ vertex with four internal lines. These types of diagrams had already occurred and been evaluated in the computation of the $\beta$-function for the theory (which of course involves only four-point vertices and therefore graphs with four or fewer internal lines). One could pursue the approach of Ref.~\cite{sann2} and use the LO and NLO semiclassical results to try to deduce lower-order (in $Q$)  perturbative results, but beyond two loops it seems (due to the increasing numbers of contributions from different powers of $Q$ involved) one would still need to supplement this information with additional perturbative computations in order to obtain the full perturbative result. 

We have also extended the analysis of Ref.~\cite{sann2} to a classically scale-invariant $U(N)\times U(N)$ theory in three dimensions. It is interesting to see how much of the formalism is rather similar. We were able to adapt perturbative results from Ref.\cite{JJ} to the present theory; we only performed the comparison with the semiclassical results up to four-loop order, but the results of Ref.~\cite{JJ} would permit a six-loop check if desired. 

In Ref.~\cite{sann3}, more general theories such as $U(N)\times U(M)$ and also more general fixed-charge operators are considered. In this case the NLO result cannot be computed analytically and so there is probably little to be gained by a detailed perturbative comparison to high orders.  It would be interesting to know if the $d=3$ NLO semiclassical contribution is equally intractable to analytic computation in these scenarios. 

We have not considered the large $\alpha_{h,y}Q$ (for $d=4$) or large $\alpha_{h,y,w}Q$ (for $d=3$) expansions in the current work, but these should be straightforward if required, following the method of Refs.~\cite{Bad2}, \cite{Bad}; though the dependence of the modes on multiple couplings might present technical problems in the curve-fitting required. 

The marriage of semi-classical methods with perturbation theory has continued to prove a productive method of probing classically scale invariant theories involving scalar field self-interactions in 3, 4 and 6 dimensions. Obviously, scale invariant theories of particular significance  that await exploration  are gauge theories, both QCD (in $d=4$) and Chern Simons theories (in $d=3$).   Also of interest would be the Thirring model (quartic fermion interactions in $d=2$). The massless Thirring model is exactly solvable inasmuch as the $nn$-points field correlation is known, which would make for an interesting test case.  Regarding the massive model, the mass spectrum of the model and the scattering matrix have been evaluated. An explicit formula for the correlations is not known; so, again, exploration from the perspective used here would be worthwhile. 

We also plan to perform a perturbative analysis of the kind we have presented here for $\phi^3$ interactions in six dimensions, where in a recent interesting article\cite{sann4} semiclassical methods have been used to check the equivalence of cubic and quartic $O(N)$ theories, as conjectured in Refs.~\cite{fei1}, \cite{fei2}.

\section*{Acknowledgements}

 We are grateful to Francesco Sannino and Chen Zhang for very helpful correspondence. DRTJ thanks the Leverhulme Trust for the award
of an Emeritus Fellowship. This research was supported by the Leverhulme Trust, STFC
and by the University of Liverpool.

\end{document}